\documentclass[useAMS,usenatbib]{mn2e}
\usepackage{graphicx}
\usepackage{times}
\usepackage{amssymb}
\usepackage{natbib}

\usepackage{pgf}
\usepackage{bm}

\def\gsim { \lower .75ex \hbox{$\sim$} \llap{\raise .27ex \hbox{$>$}} }
\def\lsim { \lower .75ex \hbox{$\sim$} \llap{\raise .27ex \hbox{$<$}} }

\begin{document}

\title[The  Magellanic Clouds]{Clues to the ``Magellanic Galaxy'' from Cosmological Simulations}

\author[Sales et al.]{
\parbox[t]{\textwidth}{
Laura V. Sales$^{1}$,     
Julio F. Navarro,$^{2}$
Andrew P. Cooper$^{1,3}$ 
Simon D. M. White$^1$,
Carlos S. Frenk$^3$ and
Amina Helmi$^4$
}
\\
\\
$^{1}$ Max Planck Institute for Astrophysics, Karl-Schwarzschild-Strasse 1, 85740 Garching, Germany\\
$^{2}$ Department of Physics and Astronomy, University of Victoria, Victoria, BC V8P 5C2,
Canada\\
$^{3}$ Department of Physics, Institute for Computational Cosmology, University of Durham, 
South Road, Durham DH1 3LE\\
$^{4}$ Kapteyn Astronomical Institute, University of Groningen, PO Box 800, 9700 AV Groningen, the Netherlands\\
%%
%\\
}

\maketitle

\begin{abstract}
  We use cosmological simulations from the Aquarius Project to study
  the orbital history of the Large Magellanic Cloud (LMC) and its
  potential association with other satellites of the Milky Way
  (MW). We search for dynamical analogs to the LMC and find a subhalo
  that matches the LMC position and velocity at either of its two most
  recent pericentric passages.  This suggests that the LMC is not
  necessarily on its first approach to the MW, provided that the
  virial mass of the Milky Way is as high as that of the parent
  Aquarius halo; $M_{200} = 1.8 \times 10^{12} \, M_\odot$. The
  simulation results yield specific predictions for the position and
  velocity of systems associated with the LMC prior to infall. If on
  first approach, most should lie close to the LMC because the
  Galactic tidal field has not yet had enough time to disperse
  them. If on second approach, the list of potential associates
  increases substantially, because of the greater sky footprint and
  velocity range of LMC-associated debris. Interestingly, our analysis
  rules out an LMC association for Draco and Ursa Minor, two of the
  dwarf spheroidals suggested by Lynden-Bell \& Lynden-Bell to form
  part of the ``Magellanic Ghostly Stream''. Our results also indicate
  that the direction of the orbital angular momentum is a powerful
  test of LMC association. This test, however, requires precise proper
  motions, which are unavailable for most MW satellites. Of the 4
  satellites with published proper motions, only the Small Magellanic
  Cloud is clearly associated with the LMC.  Taken at face value, the
  proper motions of Carina, Fornax and Sculptor rule them out as
  potential associates, but this conclusion should be revisited when
  better data become available. The dearth of satellites clearly
  associated with the Clouds might be solved by wide-field imaging
  surveys that target its surroundings, a region that may prove a fertile
  hunting ground for faint, previously unnoticed MW satellites.
\end{abstract}

\begin{keywords}
galaxies: haloes - galaxies: formation - galaxies: evolution -
galaxies: kinematics and dynamics.
\end{keywords}

%-----------------------------------

\section{Introduction}
\label{SecIntro}

The Large and Small Magellanic Clouds (LMC and SMC, respectively) are
unusual satellite galaxies. They are exceptionally bright, and so
close to the Milky Way (MW) that recent studies have concluded that
fewer than one in ten MW-like systems are expected to host satellites
with properties similar to the Clouds
\citep{Boylan-Kolchin2010,Busha2010,Lares2011,Liu2010,Guo2011,Tollerud2011}. The
short crossing time at their present Galactocentric distance suggests
that they may have already completed a number of orbits in the
Galactic potential
\citep{Murai1980,Lin1982,Gardiner1994,vanderMarel2002} while their
proximity suggests that they form a bound pair.

If the LMC/SMC are truly physically associated, then it is likely that
they were once part of a larger system: the ``Greater Magellanic
Galaxy'', to quote \citet{Lynden-Bell1982}. This idea has prompted
searches for evidence that other satellites might have been in the past
associated with the Clouds. This is encouraged, in part, by the
``polar'' distribution of the brightest Galactic satellites, which
seems to trace the orbital path of the Clouds \citep{Lynden-Bell1976}.
\citet{Lynden-Bell1995}, for example, suggested a possible association
between the LMC, SMC, Draco, Ursa Minor, Carina and Sculptor as part
of a common ``Ghostly Stream''.

The association between different satellites has recently
received renewed attention, motivated mainly by coherence in the
position and velocities of satellite pairs such as Leo IV and Leo V
\citep{Belokurov2008}) and of satellites near the Sagittarius stream;
e.g., Segue 1 \citep{Niederste-Ostholt2009}, Bootes II
\citep{Koch2009}, and Segue 2 \citep{Belokurov2009}. These ideas have
received support from the realization that most satellites should have
been accreted into the Milky Way as part of multiple systems and have
gained momentum because testing the predictions of such scenario has
become possible using realistic cosmological simulations
\citep{Sales2007b,Li_Helmi2008,Donghia2008,Ludlow2009,Klimentowski2010}.

The Magellanic Clouds are also unusual in their kinematics. The
latest proper motion measurements of stars in the Large Magellanic
Cloud \citep{Kallivayalil2006, Piatek2008} indicate that the LMC has a
much higher tangential velocity than previously thought ($V_t \sim
370$ km/s), raising questions as to whether the LMC and SMC are
actually bound to each other or even to the Galaxy as a whole.
\citet{Besla2007}, for example, have used the new kinematic data to
revise earlier orbital models \citep[see, e.g.,][]{vanderMarel2002}
and concluded that the LMC and SMC must be on the first pericentric
passage of their orbit around the Milky Way if the MW virial mass is
of the order of $\sim 10^{12}\, M_\odot$, as argued by
\citet{Klypin2002}.

The discussion above depends sensitively on the assumed virial mass of
the Galaxy. Bright satellites are more common around more massive
primaries (Guo et al. 2011), and higher primary masses make it easier to accommodate
high orbital speeds for the satellites. Even a factor of two increase
in virial mass can make a difference \citep{Piatek2008,Shattow2009},
which is certainly within the current uncertainty in virial mass
estimates for the Milky Way
\citep{Battaglia2005,Smith2007,Sales2007a,Li_White2008,Xue2008,Reid2009,Gnedin2010}.

The escape speed at $r=50$ kpc from a Navarro-Frenk-White halo with
virial mass\footnote{The virial mass, $M_{200}$, is defined as the
  mass contained within $r_{200}$, the radius of a sphere of mean
  density 200 times the critical density for closure, $\rho_{\rm
    crit}=3H^2/8\pi G$. This choice defines implicitly the virial
  radius of the halo, $r_{200}$, and its virial velocity, $V_{200}$.}
$M_{200} = 2 \times 10^{12} \, M_\odot$ is of order $\sim 500$ km/s,
which would mean that the Clouds are safely bound to the Galaxy
despite their high speed.  Further, \citet{Besla2007} show that virial
masses as high as that imply a radial period of just about 3 Gyr and
hence the possibility that the LMC and SMC have completed multiple
orbits around the Galaxy. A first infall scenario is, on the other
hand, compelling for a number of reasons, including the recent
successful modeling of the Magellanic Stream as a tidal relic of a
recent interaction between the clouds before infall into the Milky Way
\citep{Besla2010}.

The discussion above highlights the fact that basic issues such as
whether the LMC and SMC are on their first approach, or bound to each
other, or truly associated with other MW satellites, remain
unresolved. These questions are clearly interrelated. For example, the
relative positions and velocities of satellites of past LMC associates
would be very different if the Clouds are on first or second
approach. Careful modeling is therefore required to interpret current
data, especially because the orbit of the Clouds should be eroded
quickly by dynamical friction and by tidal mass loss, limiting the
applicability of models such as that of \citet{Besla2007}, which
assume that the Clouds evolve in a rigid Galactic potential.

We address these issues here by using the Aquarius Project, a series
of cosmological simulations of the formation of a Milky Way-sized halo
in the LCDM paradigm \citep{Springel2008b}. Because of their
extraordinary numerical resolution, we are able to trace the orbits
not only of massive subhalos, but also of subhalos within subhalos,
thus enabling a realistic assessment of their orbital paths before,
during, and after their accretion into the Milky Way halo. 

This paper is organized as follows.  In Sec.~\ref{SecSims} we provide
a brief description of the Aquarius simulations. We analyze the orbit
of an LMC candidate in Sec.~\ref{SecOrbit}, and the spatial and
kinematic properties of their associated subhalos in
Sec.~\ref{SecPosVel}. We use these results to explore in
Sec.~\ref{SecAssoc} which satellites of the Milky Way might have been
associated with the Clouds in the past. Sec.~\ref{SecConc} summarizes
our main conclusions.

\section{The Numerical Simulations}
\label{SecSims}

%%%%%%%%%%%%%%%%%%%%%%%%%%
\begin{center}
\begin{figure*}
\includegraphics[width=0.475\linewidth,clip]{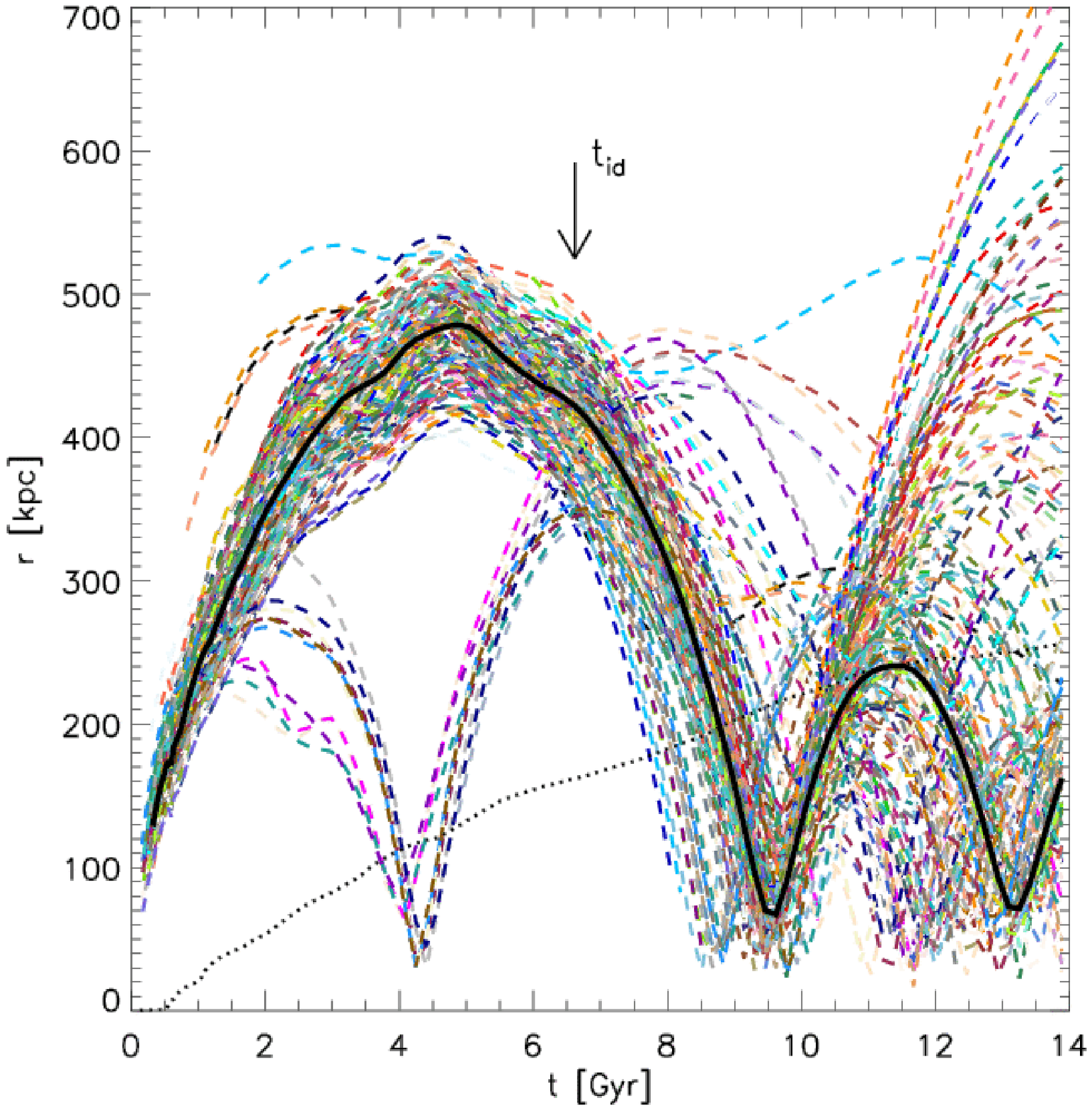}
\includegraphics[width=0.475\linewidth,clip]{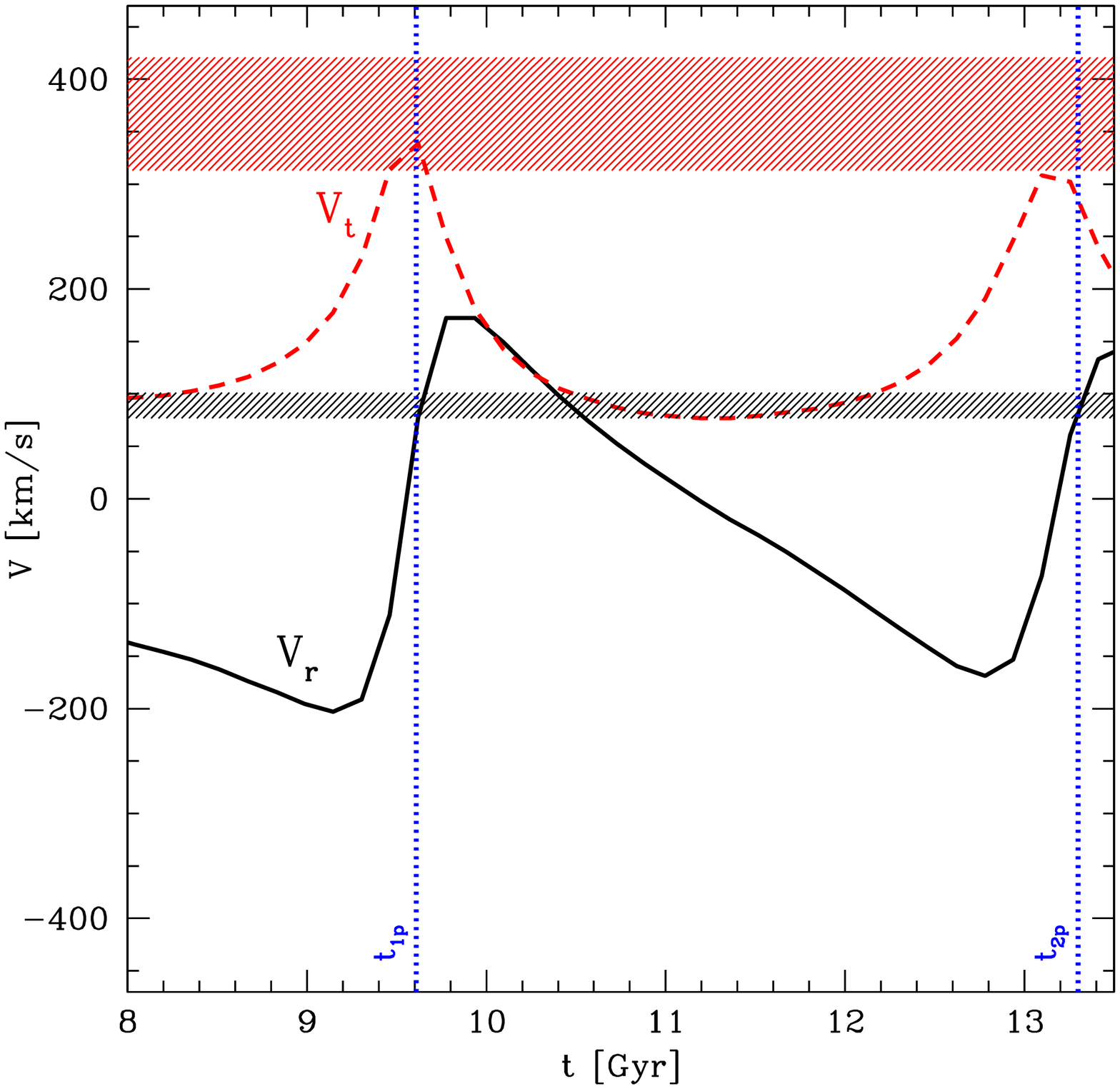}
\caption{{\it Left panel:} Time evolution of the galactocentric
  distance of the ``LMC-analog'' (LMCa) halo (solid black curve) as
  well as that of its subhalos (dashed curves), identified at $t_{\rm
    id}$ (vertical arrow).  Note the complex orbital evolution of
  associated subhalos, which include a few that are temporarily
  captured by LMCa after being ``ejected'' from the main halo during
  the tidal dissociation of a group accreted at $t\sim 4$ Gyr. The LMC
  analog enters the virial radius of the main halo (marked by the
  dotted line) at $t\sim 9$ Gyr and reaches its first pericenter at
  $t\sim 9.6$ Gyr. After first pericenter the subhalo loses $38\%$ of
  its mass and its orbital apocenter is reduced to $\sim 250$ kpc,
  about half of its first turnaround distance.  A non-negligible
  fraction of subhalos ($\sim 4\%$) are ejected into very eccentric
  orbits after first pericenter, whereas others settle onto more bound
  orbits with shorter periods. The main subhalo reaches its second
  pericenter at $t \sim 13.3$ Gyr still surrounded by some of its most
  bound companions.  {\it Right panel:} Radial and tangential velocity
  of the main subhalo as a function of time. Shaded areas correspond
  to the observed velocity of the LMC ($\pm3\sigma$) based on the
  proper motion measurements of \citet{Kallivayalil2006}.  Blue dotted
  vertical lines indicate the times $t_{1p}$ and $t_{2p}$ (near first
  and second pericenter passages) when the kinematics of LMCa best
  matches that of the LMC.}
\label{FigOrbit}
\end{figure*}
\end{center}
%%%%%%%%%%%%%%%%%%%%%%%%%%
\subsection{The Aquarius Project}

The Aquarius Project \citep{Springel2008a} is a series
of cosmological simulations of the formation of six dark matter halos
with mass consistent with that expected for the halo of the Milky Way.
The simulations assume the $\Lambda$CDM cosmology, with parameters
chosen to match the {\small WMAP} 1-year data \citep{Spergel2003}:
matter density parameter, $\Omega_{\rm M}=0.25$; cosmological constant
term, $\Omega_{\Lambda}=0.75$; power spectrum normalization,
$\sigma_8=0.9$; spectral slope, $n_s=1$; and Hubble parameter,
$h=0.73$.

The halos were identified in a large N-body simulation of a cube
137 ($100\, h^{-1}$) Mpc comoving on a side, a lower resolution version of
the Millennium-II Simulation \citep{Boylan-Kolchin2009}. This volume
was resimulated using the same power spectrum and phases of the
original simulation, but with additional high-frequency waves added to
regions encompassing the initial Lagrangian volume of each halo. The
high-resolution region was populated with low-mass particles and the
rest of the volume with particles of higher mass \citep{Power2003}.

The six Aquarius halos are labelled ``Aq-A'' through ``Aq-F''. Each
was resimulated at different resolutions in order to assess numerical
convergence. A suffix, $1$ to $5$, identifies the resolution level,
with level 1 denoting the highest resolution. Between levels 1 and 5,
the particle mass ranges from $m_p = 2 \times 10^3 \, M_\odot$ to $= 3
\times 10^6 M_\odot$. At $z=0$, the six halos have similar ``virial''
mass, roughly between $1$ and $2 \times 10^{12} M_\odot$.  For further
details of the Aquarius Project, we refer the reader to
\citet{Springel2008b} and \citet{Navarro2010}.

\subsection{Identification of the LMC analog}

We search the Aquarius simulations for accretion events that result in
systems with kinematics similar to that of the LMC. In particular,
we look for relatively massive systems (i.e., with masses exceeding
$1\%$ of the main halo mass) that are accreted relatively recently
(i.e., after $z=1$), and that have, at pericenter, distances and
velocities of order $50$ kpc and $400$ km/s, respectively.

Our best candidate is a system that accretes into the main Aq-A halo
at $z=0.51$ ($t=8.6$ Gyr, i.e., $\sim 5$ Gyr before the present
time). Just before accretion, at $z_{\rm id}=0.9$ ($t_{\rm id}=6.6$
Gyrs), this subhalo has a virial mass of $M_{200}=3.6\times 10^{10}
M_\odot$.  Our analysis below focuses on this ``LMC-analog'' halo and
its substructures associated to the same friends-of-friends group
(``LMCa group'', for short) in the Aq-A-3 simulation. This level-3
simulation has a particle mass $m_p=4.9 \times 10^{4} M_\odot$, and
therefore LMCa is resolved with more than $700,000$ particles.
SUBFIND \citep{Springel2001a} is used to identify self-bound
substructures within LMCa; at $t_{\rm id}$ there are more than $250$
LMCa subhalos with masses exceeding $1 \times 10^6 M_\odot$.

As may be seen from Fig.~\ref{FigOrbit}, the LMCa group turns around
at $t_{\rm ta}=5$ Gyr ($z=1.3$) from a distance of $r_{\rm ta}=480$
kpc (all distances and velocities quoted are physical unless
explicitly stated otherwise). LMCa reaches its first pericenter at
$t_{\rm per1}=9.5$ Gyr, with pericentric distance ($r_{\rm per1}=63$
kpc) and speed ($V_{\rm per1}=355$ km/s) comparable to those of the
LMC. The LMCa orbit becomes substantially more bound after first
pericenter due to dynamical friction and tidal mass loss, so that it
reaches a distance of only $240$ kpc at its second apocenter, at
$t=11.5$ Gyr. Its radial period reduced to $3.8$ Gyr, it goes
through pericenter again at $t_{\rm per2}=13.3$ Gyr, with similar
pericentric distance and speed as the first (see right panel of
Fig.~\ref{FigOrbit}).

The best match to the kinematic properties of the LMC occurs just
after each of these pericentric passages, at times that we will denote
$t_{1p}$ and $t_{2p}$. These are shown in the right panel of
Fig.~\ref{FigOrbit} with vertical dotted lines. At these times, the
virial mass of the host halo is $1.6$ and $1.8 \times 10^{12} \,
M_\odot$, respectively. At $t = t_{1p} (t_{2p})$, the satellite is at
a distance of $65 (69)$ kpc, with radial velocity $V_r=78 (89)$ km/s
and tangential velocity $V_t=347 (302)$ km/s. In the analysis that
follows we focus on the properties of LMCa and its associated subhalos
at $t_{1p}$ and $t_{2p}$.

  These values are in reasonable agreement with the latest LMC
  measurements; in particular, the tangential speed at first
  pericenter agrees within $1\sigma$ with the results of
  \citet{Kallivayalil2006} and of \citet{Piatek2008}. The velocities
  are lower by $\sim 15\%$ at second pericenter, mainly because the
  apocenter of the orbit has been reduced to $230$ kpc from the $480$ kpc
  reached at turnaround. This argument would appear to
  favour a first-pericentric interpretation for the LMC orbit, in tune
  with recent suggestions (see, e.g., Besla et al. 2007, Boylan-Kolchin et al.
  2010, Busha et al. 2010b, Tollerud et al. 2011). We
  urge caution, however, with this interpretation. The actual
  pericentric speed will depend sensitively on the actual turnaround
  radius and on the mass of the Galaxy, both of which are rather
  uncertain. Further, we do not expect an exact match between LMC and
  LMCa because of the limited statistics that the six Aquarius halos
  allows. A more robust result seems to be that, because of the large
  pericentric distance, systems analogous to the LMC should experience
  a relatively moderate loss of orbital energy, allowing them to
  return to its second pericenter with similar speed to the first.  If
  LMCa had had a slightly larger first-pericenter speed then quite
  possibly the second pericenter would have been in better agreement
  with the LMC than the first. We conclude that multiple passages cannot
  be confidently ruled out by this evidence alone.
 \nocite{Besla2007,Boylan-Kolchin2010, Busha2010b,Tollerud2011}

%%%%%%%%%%%%%%%%%%%%%%%%%%
\begin{center}
\begin{figure*}
\includegraphics[width=0.475\linewidth,clip]{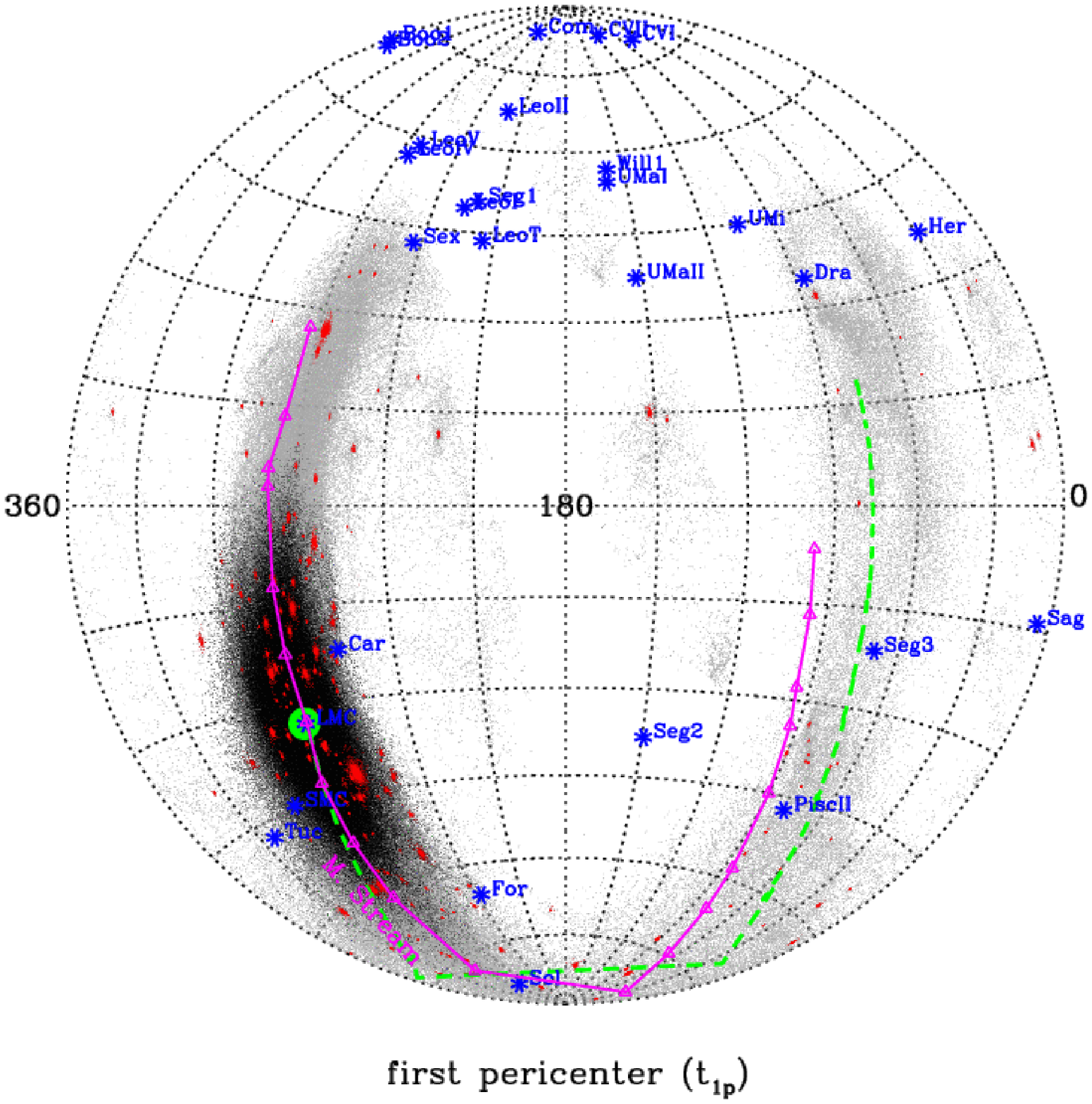}
\includegraphics[width=0.475\linewidth,clip]{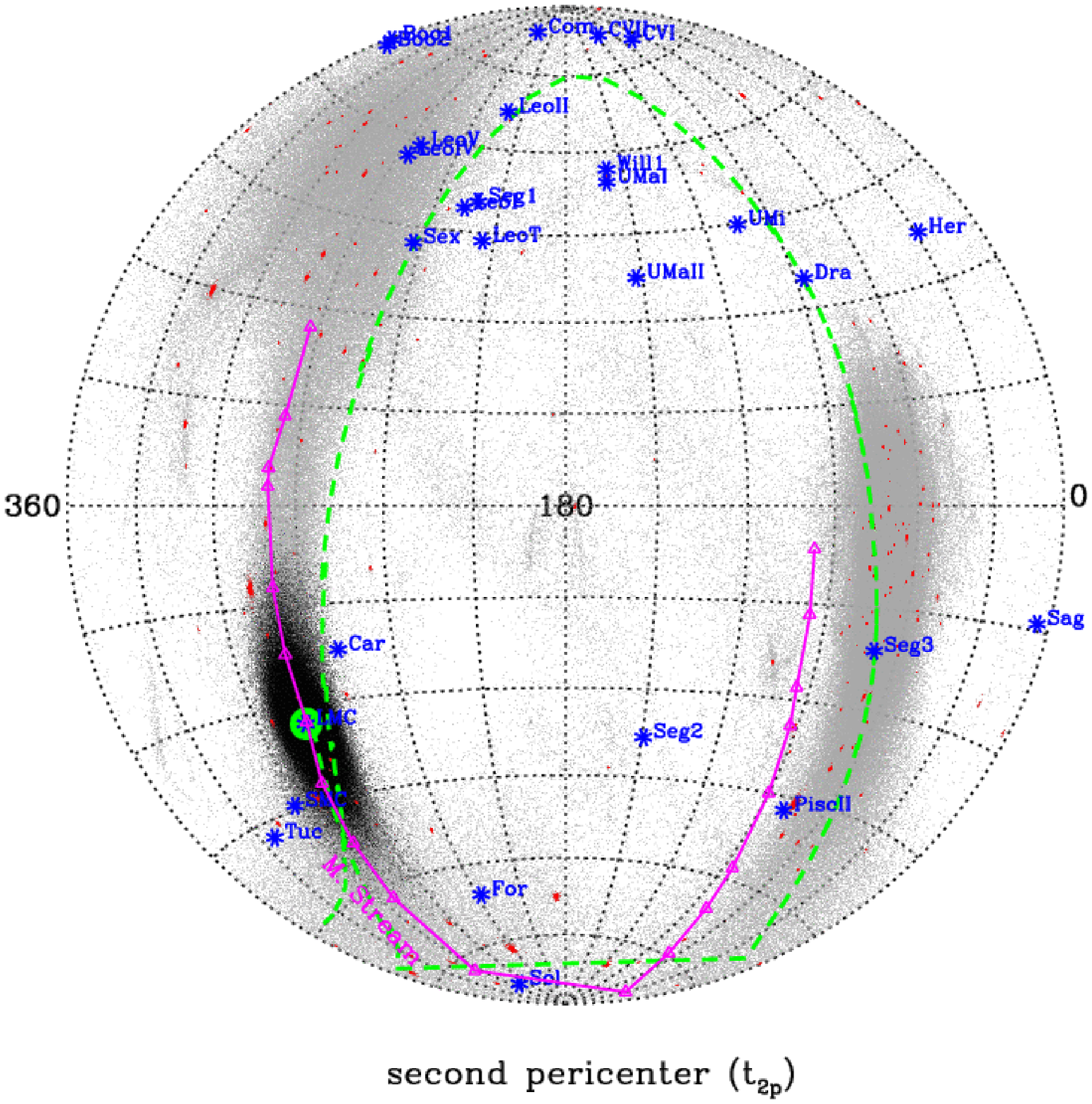}
\caption{Aitoff projection of all particles belonging to the LMCa
  group at two different times, $t_{1p}$ (near 1st pericenter, left
  panel) and $t_{2p}$ (near 2nd pericenter, right panel), when the
  kinematics of the most massive LMCa member (shown by a green circle)
  matches best that of the Large Magellanic Cloud (see
  Fig.~\ref{FigOrbit}). Dots denote all individual particles
  within the virial radius of the LMCa at the time of identification, $t_{\rm
    id}$. Those colored black are still bound to LMCa; grey are those
  that have become unbound as a result of the tidal interaction with
  the primary halo. Particles in red are those belonging to self-bound
  substructures of LMCa. The positions in the sky of all known Milky
  Way satellites are also shown and labelled in blue. The orbital path
  of LMCa is shown by the dashed green curve; a magenta curve traces
  the location of the Magellanic Stream, taken from
  \citet{Nidever2010}. The reference frame of the simulation has been
  rotated so that the position in the sky and the proper motion of
  LMCa match those of the LMC. Note how all particles and
  substructures associated with LMCa stretch along the orbital
  path. At $t_{1p}$ (panel on the left) most associated subhalos lie
  near the LMC since tidal stripping has not yet had enough time to
  disrupt the system. At $t_{2p}$ (panel on the right) the particles
  and subhalos associated with LMCa are more widely spread across the
  sky but still trace the orbital path. The LMCa footprint can be
  used, together with velocity information, to investigate whether
  other Milky Way satellites have been associated with the LMC in the
  past. See text for further details.}
\label{FigAitoff}
\end{figure*}
\end{center}
%%%%%%%%%%%%%%%%%%%%%%%%%%

\section{Results}
\label{SecResults}

\subsection{The Orbit of the LMCa Group}
\label{SecOrbit}

The left panel of Fig.~\ref{FigOrbit} shows the radial evolution of
LMCa and its $250$ most massive associated subhalos (identified at
$t_{\rm id}$ within the same friends-of-friends group). This shows the
complex orbital behaviour of LMCa subhalos as the group gets disrupted
in the tidal field of the primary halo. As emphasized by
\citet{Sales2007b} and \citet{Ludlow2009}, many satellites undergo
drastic changes in energy and angular momentum during pericentric
passage, causing some systems to achieve escape velocity and to leave
the primary halo altogether.

Interestingly, the LMCa group itself contains captured ``escapees''
from another group. As may be seen in the left panel of
Fig.~\ref{FigOrbit}, a number of subhalos belonging to LMCa at $t_{\rm
  id}$ were actually accreted much earlier into the main halo and
ejected from it when their parent group was disrupted at $t\sim 4$
Gyr. By chance these were then temporarily captured by LMCa at $t_{\rm
  id}$ before plunging again into the main halo.

These changes limit the applicability of the usual procedure of
looking for satellite associations by searching for objects of common
energy and/or angular momentum \citep[see, e.g.,][]{Lynden-Bell1995}.
Instead of making simple assumptions we shall use directly the 6D
information from the simulations to assess the probability that
other Milky Way satellites are associated with the LMC.

\subsection{Position and Kinematics of the LMCa Group}
\label{SecPosVel}

We plot in Fig.~\ref{FigAitoff} the position in the sky (in an Aitoff
projection) of all particles identified at $t_{\rm id}$ to belong to
LMCa. The panel on the left corresponds to first pericenter, $t_{1p}$,
and that on the right to second pericenter, $t_{2p}$. The simulation
reference frame has been rotated so that the position in the sky and
the direction of motion of the main LMCa subhalo (indicated by a green
circle) coincide with those of the LMC. The orbital path is shown by a
green dashed line; the line in magenta traces the Magellanic Stream
\citep{Nidever2010}. Particles in black are those currently still
bound to LMCa, those in grey correspond to tidally stripped LMCa
material. Particles in red highlight the LMCa substructures that
remain self-bound.

%%%%%%%%%%%%%%%%%%%%%%%%%%
\begin{center}
\begin{figure*}
\includegraphics[width=0.473\linewidth,clip]{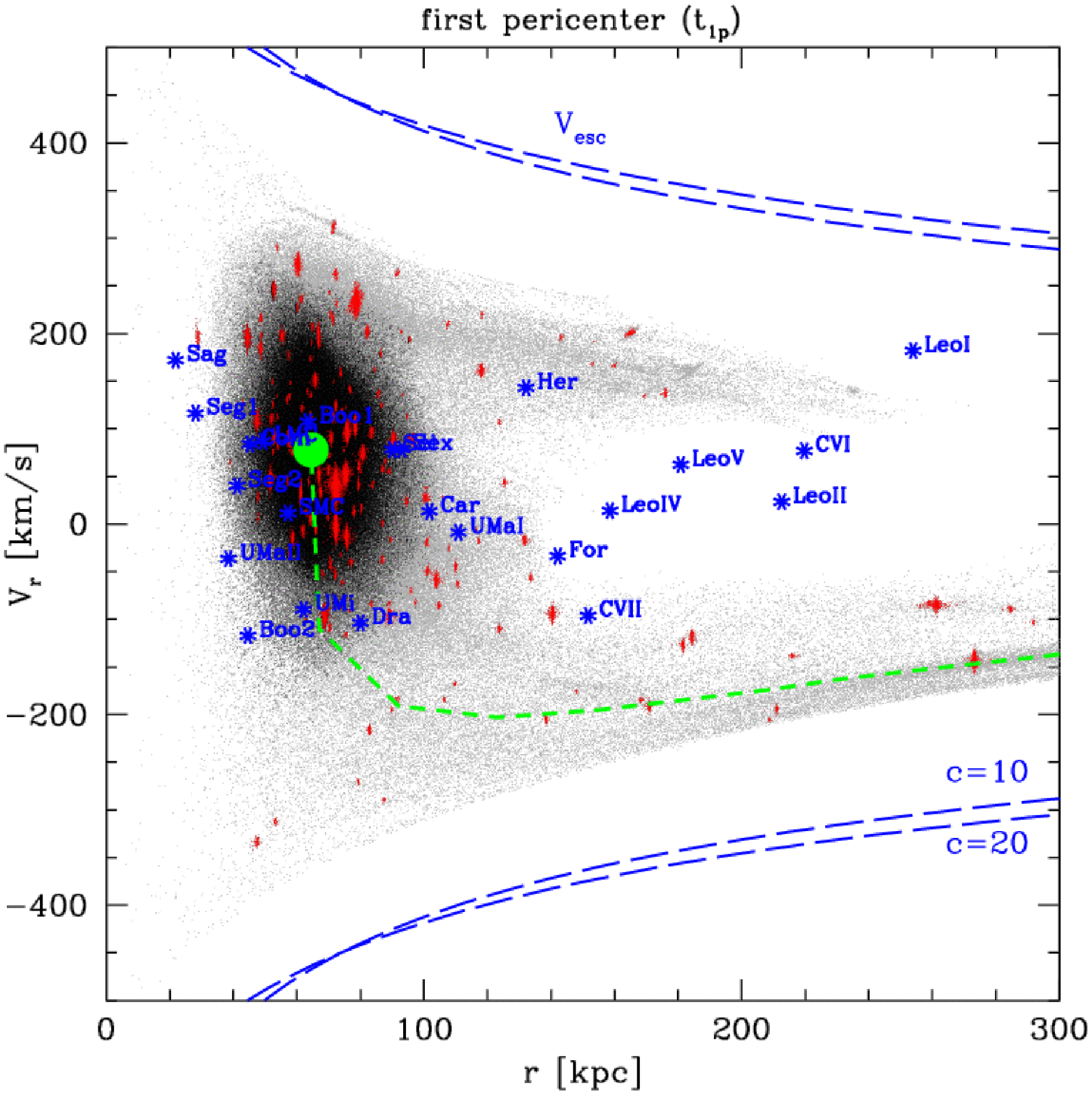}
\includegraphics[width=0.473\linewidth,clip]{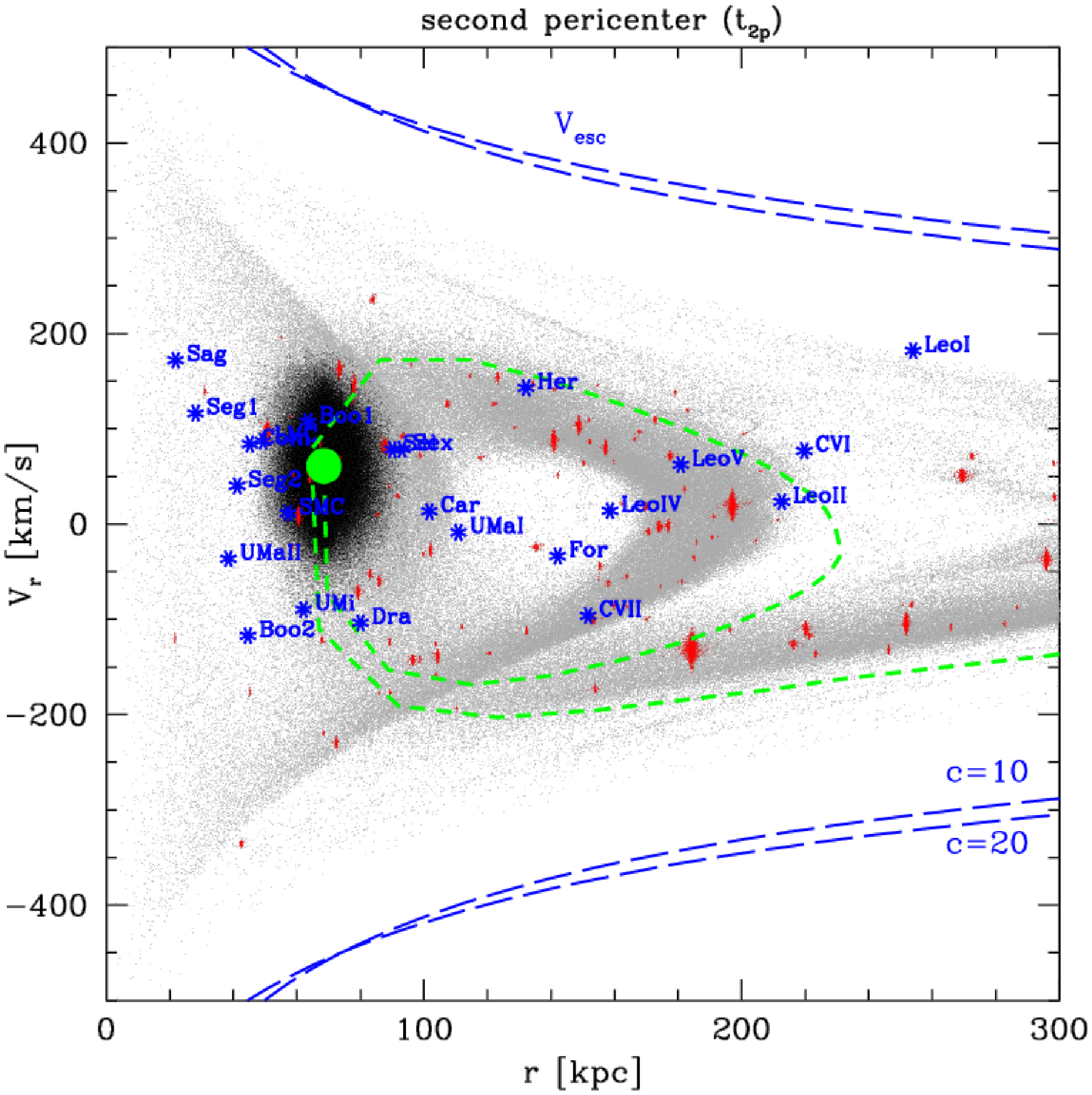}
\caption{Galactocentric distance versus radial velocity for all
  particles in the LMCa group, compared with those of known Milky Way
  dwarfs. Color coding is the same as in Fig.~\ref{FigAitoff}.  Left
  and right panels correspond to $t=t_{1p}$ and $t=t_{2p}$,
  respectively.  For reference, the local escape velocity of a
  Navarro-Frenk-White halo \citep{Navarro1996,Navarro1997} is indicated by the blue long-dashed
  curves, for two values of the concentration, $c=10$ and $20$.  Note
  that the tidal debris from LMCa is constrained to a well defined
  region in the ($r,V_r$) plane, a fact that may be used to gauge the
  likelihood of association with the LMC of individual dwarfs.}
  \label{FigRVr}
\end{figure*}
\end{center}
%%%%%%%%%%%%%%%%%%%%%%%%%%

Fig.~\ref{FigAitoff} shows the complex footprint on the sky traced by
LMCa and its debris, and how it changes substantially from first to
second pericentric passage. At first approach (left panel of
Fig.~\ref{FigAitoff}), most LMCa subhalos cluster around the LMC
because tidal stripping has not yet had enough time to disrupt the
system. Nevertheless, there are a few tidally-stripped systems (those
more loosely bound to LMCa at $t_{\rm id}$), and they roughly trace the
orbit of LMCa along a nearly polar circle. The sky is not uniformly
populated by the stream; for example, because the pericentric passage
occurs near the Southern Galactic Pole very few LMCa subhalos have a
chance of reaching the northern Galactic cap by $t_{1p}$. Some systems
stray from the orbital plane despite their previous association, but
these cases are rare and correspond to subhalos populating the outskirts
of the LMCa group at $t_{\rm id}$.

The orbital path is still discernible in the tidal stream at second
pericenter, but the debris populates now both hemispheres in the sky.
%$360^\circ$ of the nearly polar circle.
At this time LMCa has lost more than $70\%$ of its original
mass to tides, compared with only $40\%$ at $t=t_{1p}$. The distribution
of the debris on the sky would certainly be wider for a progenitor of
larger mass than the one considered here

Blue asterisks in Fig.~\ref{FigAitoff} indicate the position of all
known Milky Way satellites, including the ultra-faint dwarfs
discovered by SDSS, which are located mainly in the northern Galactic
cap (see Table ~\ref{TabMWsats} for the compilation of values used). 
Several dwarfs lie along the debris path and therefore could
potentially have been associated with the LMC. These include, at
$t=t_{1p}$, Sextans (Sex), Draco (Dra), Segue 3 (Seg3), Pisces II
(PiscII), Sculptor (Scl), Fornax (For), Tucana (Tuc), Carina (Car),
Leo II, and the SMC. Because the debris is more spread out at
$t_{2p}$, the list includes then further dwarfs, such as those in the
constellation of Leo.

Of course, true association requires not only coincidence with the
LMCa tidal stream in projection, but also in distance and velocity. We
investigate this in Fig.~\ref{FigRVr}, where we plot the
Galactocentric distance and radial velocity of LMCa particles for
$t=t_{1p}$ (left) and $t=t_{2p}$ (right). Color coding is the same as
in Fig.~\ref{FigAitoff}. Data for all known Milky Way satellites 
(as listed in Table \ref{TabMWsats}) are also shown in
each panel, for comparison. Since the LMCa debris is
confined to specific regions in the ($r,V_r$) plane, these data may be
used to test the association of any individual dwarf with the LMC,
subject to assuming that the LMC is either on its first or second
approach to the Galaxy. We explore these associations next.

\subsection{Association with the Clouds}
\label{SecAssoc}

To be deemed an ``LMC associate'' a satellite must satisfy at least the
following three conditions: (i) it should fall in the celestial sphere
within the footprint of the LMCa group; (ii) it should be at a
Galactocentric distance consistent with that of LMCa particles at the
same location in the sky; and (iii) it should have Galactocentric
radial velocity also consistent with LMCa in the same region. In
principle a further condition involving the tangential velocity could
be added to the list, but the preliminary nature of most proper motion
estimates for Milky Way satellites implies that the results are unlikely to
be conclusive at this stage (see the Appendix for more details). 
We discuss conditions (i) to (iii) below.

\subsubsection{Sky proximity}
\label{ssec:skyprox}

We can quantify the proximity in the sky of any dwarf to the LMCa
stream by computing the angular distance to the n$^{\rm th}$ nearest
LMCa particle in projection and comparing that with the probability of
obtaining a similar distance (or smaller) for a random point in the
sky. Note that the analysis that follows uses all LMCa particles
  (rather than just subhalos) when comparing with the Milky Way
  satellites, since this provides a more complete sampling of the
  distribution in phase space expected for LMC debris.

The histograms in Fig.~\ref{FigHisto} show the distribution of
the angular distance to the 50th nearest LMCa particle, $\delta_{50}$,
computed for 10,000 random points in the celestial sphere; the median
of the distribution is at $\delta_{\rm lim}=4^\circ$ and $\delta_{\rm
  lim}=2^\circ$ for the first and second pericenter passages,
respectively. As may be seen from the inserts in Fig.~\ref{FigHisto}
the footprint of the LMCa stream is traced faithfully by regions
satisfying $\delta_{50}<\delta_{\rm lim}$ (shown in red) and we shall
use this condition to decide which dwarfs are likely to be associated
with the LMC.

According to Fig.~\ref{FigHisto}, at first pericenter only $8$
dwarfs satisfy the proximity condition specified above but, because
the LMCa footprint covers a much larger fraction of the sky at second
pericenter, $17$ dwarfs pass this constraint then. These numbers are
reasonably insensitive to our choice of n=50, especially at first
pericenter; choosing the 100th nearest neighbour results in $9$
likely-associated dwarfs at $t_{1p}$ and $22$ at $t_{2p}$.

\subsubsection{Radial velocity}
\label{ssec:vr}

%%%%%%%%%%%%%%%%%%%%%%%%%%
\begin{center}
\begin{figure*}
\includegraphics[width=0.475\linewidth,clip]{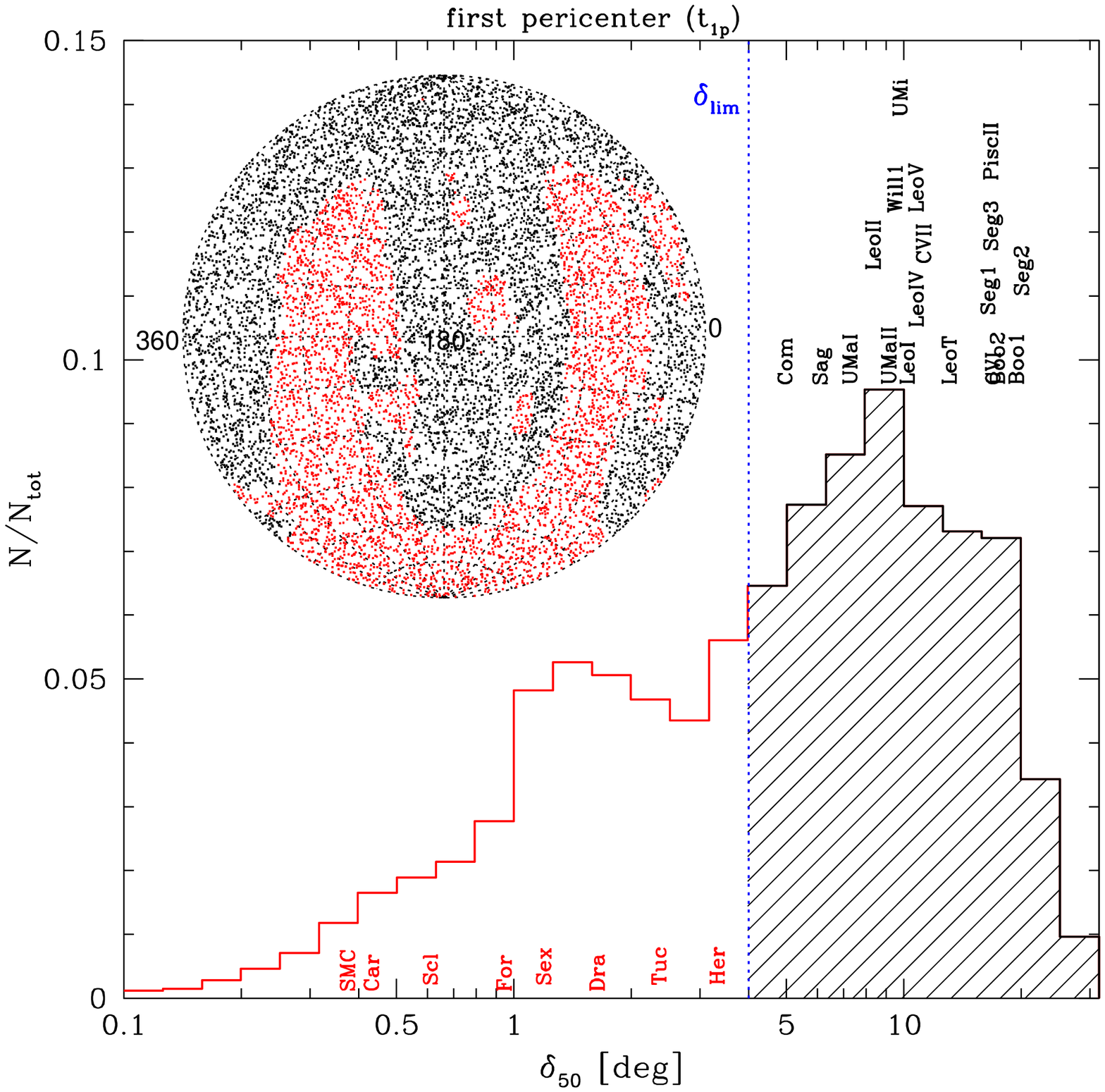}
\includegraphics[width=0.475\linewidth,clip]{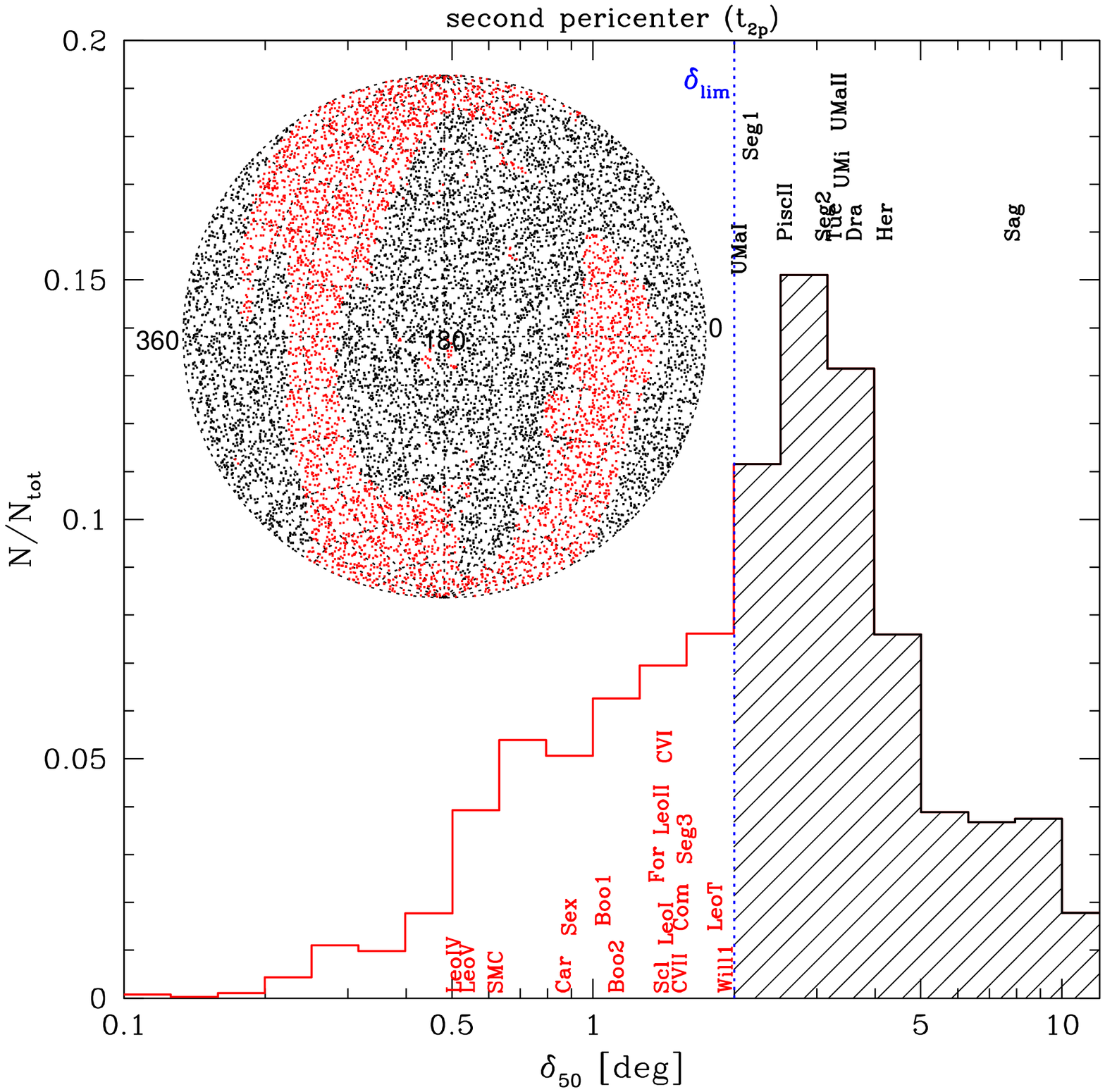}
\caption{Distribution function of $\delta_{50}$, the angular size
  (radius) of the circle that contains the $50$ nearest LMCa particles
  in projection, computed for $10,000$ random points in the celestial
  sphere. Left and right panels correspond to $t=t_{1p}$ and
  $t=t_{2p}$, respectively.  $\delta_{50}$ is a useful measure of
  proximity to the LMCa stream in the sky for any given point. In the
  inserted Aitoff maps, red is used to highlight points with
  $\delta_{50}<\delta_{\rm lim}$, where $\delta_{\rm lim}$ is the
  median of the random distribution. Black points correspond to
  $\delta_{50}>\delta_{\rm lim}$. As demonstrated by the Aitoff
  inserts, $\delta_{50}$ is a useful measure of association with the
  stream. We show in each panel the values of $\delta_{50}$
  corresponding to each Milky Way dwarf, and retain only those with
  $\delta_{50}<\delta_{\rm lim}$ as potentially associated with the
  LMC.}
\label{FigHisto}
\end{figure*}
\end{center}
%%%%%%%%%%%%%%%%%%%%%%%%%%

%%%%%%%%%%%%%%%%%%%%%%%%%%
\begin{center}
\begin{figure*}
\includegraphics[width=0.75\linewidth,clip]{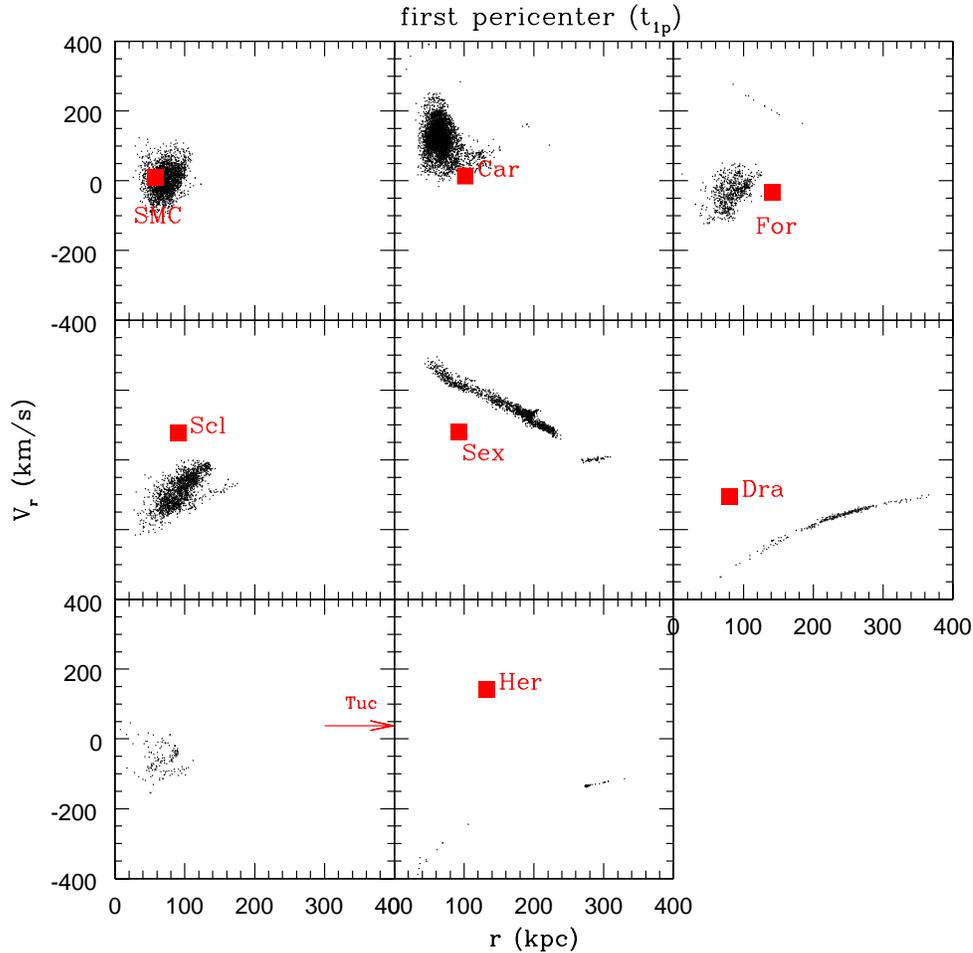}
\caption{Galactocentric radial velocity and distance for dwarfs
  (filled squares) deemed possibly associated with the LMC according
  to the criterion of Fig.~\ref{FigHisto}; i.e., $\delta_{50}<\delta_{\rm
    lim}$ at $t=t_{1p}$. Each panel also shows the $r$ and $V_r$ of
  all LMCa particles (dots) within a circle of radius $\delta_{\rm
    lim}$ centered at the position of each dwarf. This probes
  graphically whether the positional association indicated by
  proximity to the stream in the sky is corroborated by the velocity
  data. This test indicates that the SMC is the only known satellite
  clearly associated with the LMC if the Clouds are on their first
  pericentric approach. Aside from the SMC only Carina and Fornax seem
  marginally consistent with an LMC association. On the other hand,
  this test seems to rule out a possible association for all other candidates.}
\label{FigRVrSat1p}
\end{figure*}
\end{center}
%%%%%%%%%%%%%%%%%%%%%%%%%%

\begin{center}
\begin{figure*}
\includegraphics[width=0.75\linewidth,clip]{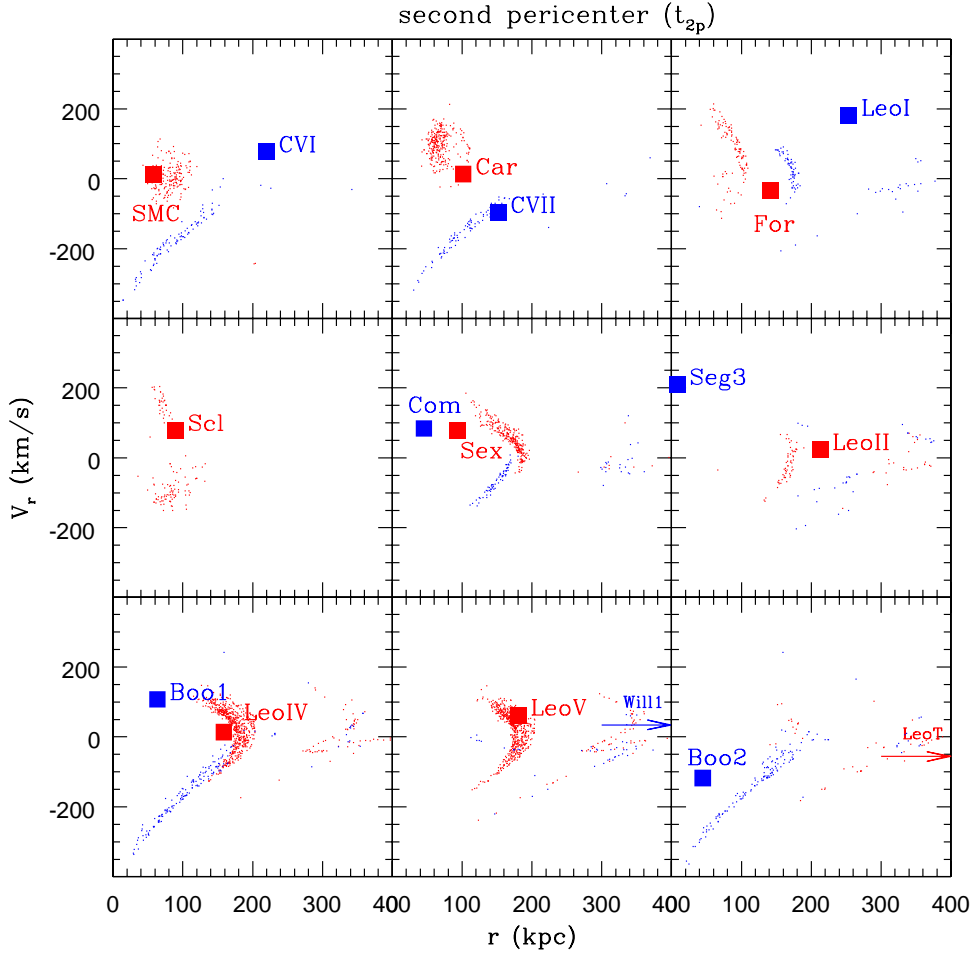}
\caption{Same as Fig.~\ref{FigRVrSat1p} but for the second pericenter
  passage. Because the stream covers a wider area in the sky, the
  number of candidate dwarfs increases. We show {\it two} dwarfs per
  panel; arrows indicate systems that lie beyond the plotted
  region. In each panel red (blue) dots are stream particles in the
  $\delta_{\rm lim}$ circle around the dwarf indicated by a red
  (blue) square. This exercise confirms the
  association of the SMC, and suggests that several other dwarfs have
  kinematics consistent with association with the LMC. See text for
  further discussion.}
\label{FigRVrSat2p}
\end{figure*}
\end{center}

Figs.~\ref{FigRVrSat1p} and ~\ref{FigRVrSat2p} are then used to check
which of these candidate dwarfs are also associated with the stream in
distance and velocity. The various panels in these figures show the
Galactocentric distance and radial velocity of each of the dwarfs in
the LMCa sky footprint, together with those of all LMCa particles
closer than $\delta_{\rm lim}$ in the sky from each dwarf. This
enables an intuitive test of which satellites have distances and
velocities consistent with LMC association.

For example, the top left panel of Fig.~\ref{FigRVrSat1p} corresponds
to the SMC and it shows that there are plenty of LMCa particles with
distances and radial velocities coincident with the SMC at first
approach, endorsing a true association between the LMC and the SMC. An
opposite example is provided by Draco; the data in
Fig.~\ref{FigRVrSat1p} show that, although it is within the LMCa
footprint, all of the particles associated with LMCa at Draco's
location have discrepant velocities and/or distances. This is because
Draco is projected onto the trailing stream at first pericenter, and
one would expect then very large negative radial velocities at Draco's
Galactocentric distance. A similar analysis may be carried out for
each dwarf individually, but it should be clear from
Fig.~\ref{FigRVrSat1p} that, aside from the SMC, only Carina and
Fornax seem to have a reasonable chance of being associated with the
Clouds if they are on first approach.

The situation is less clear-cut at second pericenter, because the LMCa
footprint is larger, and because the stream now has multiple wraps,
which allows for a wider range of velocities in a given direction in
the sky to be consistent with LMCa association. The right panel of
Fig.~\ref{FigHisto} indicates that $17$ dwarfs pass the sky proximity
test; these are shown (two per panel except in the middle left) in
Fig.~\ref{FigRVrSat2p}. Inspection of each panel shows that, aside
from the SMC, Carina, Canes Venatici II, Sculptor, Leo II, Leo IV, and
Leo V could in principle be associated with the LMC. The situation of
Fornax, Canes Venatici I, and Sextans is less clear but one would be
hard pressed to rule out an association in such cases. Bootes (1 and
2), Segue 3, LeoT, Willman I and Coma are all unlikely to be
associated with the LMC.

%%%%%%%%%%%%%%%%%%%%%%%%%%
%%%%%%%%%%%%%%%%%%%%%%%%%%%%%%

\subsubsection{Orbital angular momentum}
\label{ssec:angmom}

The sky proximity and radial velocity constraints discussed above are
sensitive to the fact that we consider here a single LMC look-alike
system. This is compounded by the relatively low mass of LMCa,
compared with the $1.3\times 10^{11} M_\odot$ suggested by the
abundance-matching analysis of \citet{Boylan-Kolchin2010}. A more
massive subhalo may leave a broader footprint on the sky and lead to a
wider range of radial velocities consistent with LMC-associated
debris. More conclusive statements about the likelihood of association
of the candidate satellites identified in the previous subsections
require accurate measurements of the proper motion in order to
constrain their tangential velocity and to verify that they follow
orbits roughly aligned with the orbital plane of the LMC. Indeed, as
we show below, the direction of the orbital angular momentum of a
dwarf might be one of the cleanest tests of association with the
Clouds.

\begin{center}
\begin{figure*}
\includegraphics[width=0.475\linewidth,clip]{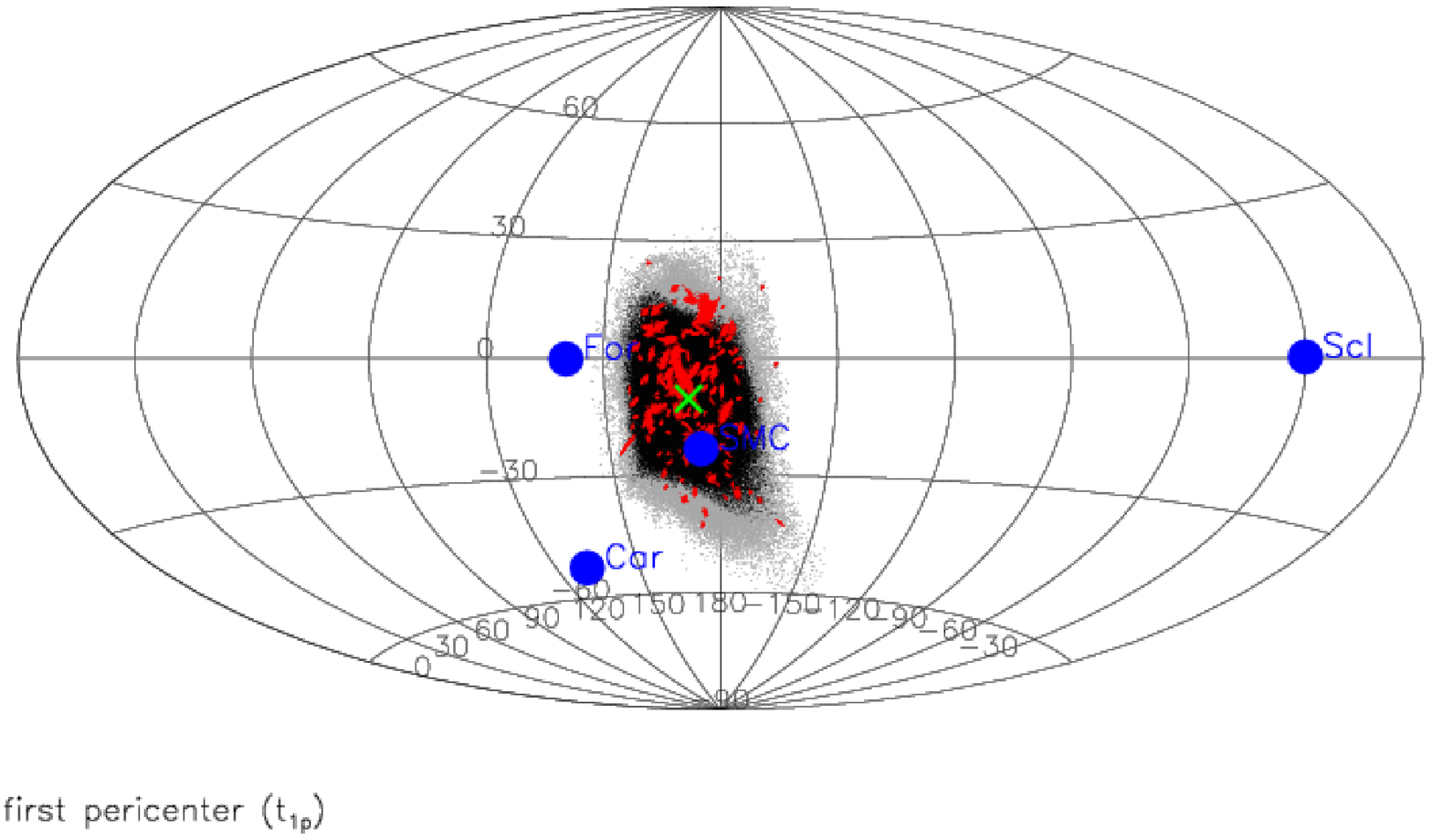}
\includegraphics[width=0.475\linewidth,clip]{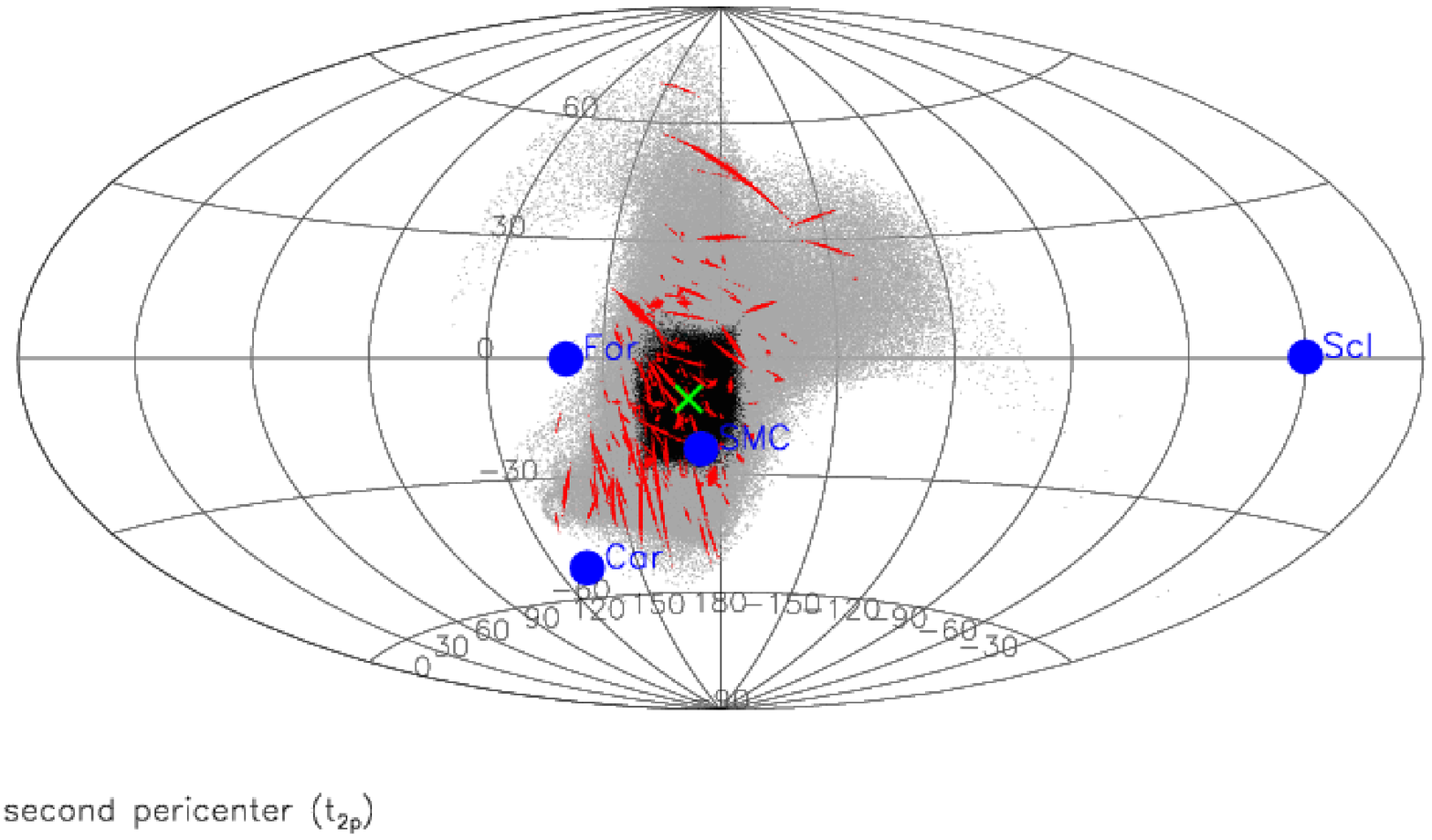}
\caption{Direction of the orbital angular momentum of LMCa at first
  and second pericenter passages, compared with those of dwarfs with
  published proper motions  (SMC, Fornax, Sculptor and
    Carina). The directions are shown in an Aitoff projection of
  Galactocentric coordinates ($l_G$ and $b_G$). The color coding is
  the same as in Figs.~\ref{FigAitoff} and ~\ref{FigRVr},  with
    the green cross indicating the orbital pole of the central subhalo
    in LMCa. Note that the viewing perspective is different
    from that in Fig.~\ref{FigAitoff}, $l >180^\circ$ in the former
    correspond to negative longitudes here.
  All satellites associated with the LMC before infall are expected to
  have their orbital poles in a well defined region of the sky
  centered approximately at $(l_G,b_G) \sim (170^\circ,-10^\circ)$. Of
  the 4 dwarfs with measured full spatial motions only the SMC seems
  obviously associated with the LMC.}
\label{FigPoles}
\end{figure*}
\end{center}

This is shown in Fig.~\ref{FigPoles}, where we show, in an Aitoff
projection of Galactocentric coordinates $(l_G,b_G)$, 
the direction of the orbital angular momentum of LMCa and
its associated substructures at first and second pericentric
passage. Because of the nearly polar orbit the direction of the
angular momentum of most LMCa-associated material is roughly on the
Galactic plane, pointing in the direction of the Sun from the Galactic
center; i.e., $l_G=180^\circ$, $b_G=0^\circ$. This tight alignment is
preserved at second approach, albeit with larger scatter.

In Galactocentric Cartesian coordinates, with the $X$-axis pointing
away from the Sun, $Y$-axis defined positive in the direction of
Galactic rotation, and positive $Z$-axis in the direction of the
Galactic North Pole, this implies that the $X$-component of the
orbital angular momentum ($j_X$) of associated satellites should be
negative and much larger in magnitude than $j_Y$ or $j_Z$. This may be
seen in Table~\ref{TabPropMot1}, which lists the components of the
{\it unit} vector identifying the direction of the (average) angular
momentum of particles associated in the sky with the candidate dwarfs
identified in the previous subsection. Note that, with no exception,
the angular momentum points clearly toward $-X$, the anti-Galactic
center direction.

This result makes strong predictions regarding the tangential velocity
of the candidate satellites, which can be checked against observation
for the few satellites with available proper motions (SMC:
Kallivayalil (2006b), Carina: \citet{Piatek2003}, Fornax:
\citet{Piatek2007} and Sculptor: \citet{Piatek2006}). Inspection of
Table~\ref{TabPropMot1} and Fig.~\ref{FigPoles} shows that, of the
four satellites with published spatial velocities, only the SMC
appears associated with the LMC.  None of the other three (Carina,
Fornax, and Sculptor) seems obviously associated with the Clouds
according to this test (see last three columns of Table
  \ref{TabMWsats}).  In hindsight this not entirely surprising,
  given the very dissimilar chemical enrichment patterns and gas
  content of the LMC and SMC compared with other Galactic satellites
  (Mateo 1998, Carrera et al. 2008, Harris et al. 2009, Kirby et al. 2011a,b). Explaining
    what drives the diversity in star formation history, metal
    enrichment, and gas fractions of Galactic satellites remains a
    prime challenge for dwarf galaxy formation models.
We hasten to add, however, that, as the recent revision to the proper
motion of the LMC illustrates \citep{Kallivayalil2006}, proper motion
measurements are exceedingly difficult, and hence our conclusion
should be revisited when new, more accurate data become
available. 
\nocite{Mateo1998,Carrera2008,Harris2009,Kirby2011a,Kirby2011b}

\section{Summary and Conclusions}
\label{SecConc}

We use cosmological N-body simulations from the Aquarius Project to
study the orbit of the LMC and its possible association with the SMC
and other Milky Way satellites in light of new proper motion data
\citep{Kallivayalil2006, Piatek2008}. We search the simulations for
LMC dynamical analogs; i.e., accreted subhalos with pericentric
distance ($\sim 50$ kpc) and velocity ($\sim 400$ km/s) matching those
of the Clouds.

One suitable candidate (LMCa) is a $3.6 \times 10^{10} M_\odot$ system
accreted at $z\sim 0.5$ ($t=8.7$ Gyr) by Aq-A, a halo that, at $z=0$
has a virial mass of $1.8 \times 10^{12} M_\odot$. LMCa turns around
from a distance of $480$ kpc at $t_{\rm ta}\sim 5$ Gyr ($z=1.3$),
accretes into the Milky Way halo at $t=8.6$ Gyr ($z=0.5$), and
completes two pericentric passages by $z=0$. 

We use the positions and velocities of particles belonging to LMCa
before infall in order to trace the orbital evolution of
LMC-associated satellites and to inform the analysis of the likelihood
that other Milky Way satellites were accreted in association with the
Magellanic Clouds. Our main conclusions may be summarized as follows.

Near each pericentric passage the kinematic properties of LMCa match
approximately those of the LMC. This implies that (i) the orbit of the
LMC is not particularly unusual given the halo virial mass, and that
(ii) it is difficult to decide, using only kinematical data, whether
the LMC is on first approach or has already completed a full orbit.

If the LMC is {\it on first approach}, then most of its associated
subhalos should be tightly clustered around its location. Although
rare, some LMC-associated systems may still be found well away from
the LMC but along the orbital path of the group. Since none of them
has completed a single orbit there are strong position- radial
velocity correlations that may be used to identify which satellites
might have been accreted together with the LMC. 

Of the known Milky Way satellites only the SMC is clearly associated
with the LMC. A case can also be made for Fornax, Carina, and
Sculptor, but it is not a particularly compelling one. This is
specially true when considering available proper motion data, which
suggest that the orbital planes of these three satellites are not
aligned with that of the Clouds.

If the LMC is near its {\it second pericenter} then several further
dwarfs qualify for association. Leo II, Leo IV and Leo V, in
particular, show strong spatial and velocity coincidence with the
tidal debris from LMCa, making them prime candidates for past
association with the LMC. Persuasive, but hardly conclusive, cases can
also be made for a handful of other dwarfs, such as Canes Venatici II
and Leo I.  These tentative associations may be firmed up or refuted
using the full spatial velocity, for which our simulations make a
strong prediction: it must be such that the direction of the orbital
angular momentum should point unambiguously in the anti-Galactic
center direction.

We expect then few, if any, known Milky Way satellites to be
associated with the Clouds, especially if they are on {\it first
  approach}.  The simulations, however, make a very clear prediction
in that case: most satellites associated with the Clouds have yet to
disperse and should therefore be very near them. How many could we
expect? The number depends strongly on the virial mass of the Clouds
and is, therefore, highly uncertain. If the mass is as high as $1.3
\times 10^{11} M_\odot$, as suggested by the abundance-matching
analysis of \citet{Boylan-Kolchin2010}, then we would expect it to
have of order $7$ subhalos with peak circular velocities exceeding
$20$ km/s \citep{Springel2008b}. This, according to the recent model
of \citet{Font2011}, is the minimum halo potential depth required 
to host a luminous dwarf. 

Although the numbers seem modest, one may be encouraged by the fact
that the LMC does have at least one companion (the SMC), so the
possibility that they are part of a larger ``Magellanic Galaxy''
should not be dismissed too quickly.  Surveys designed to target
  the sky around the LMC (SkyMapper\footnote{http://www.mso.anu.edu.au/skymapper/},
  MAPS \citep{Nidever2011}, NOAO Outer Limits Survey \citep{Saha2010})
  should help to unravel the history of the LMC and its companions.  The
surroundings of the Clouds might be hiding a trove of new Milky Way
satellites awaiting discovery.

%%%%%%%%%%%%%%%%%%%%%%%%%%
%%%%%%%%%%%%%%%%%%%%%%%%%%

%TABLE%%%%%%%%%
\begin{center}
\begin{table*} 
  \caption{Cartesian components of the unit vector that characterizes the
    direction of the (average) angular momentum of LMCa particles near
    each candidate LMC-associated satellite,
    according to the discussion of Fig.~\ref{FigRVrSat1p} and Fig.~\ref{FigRVrSat2p}.
    For each dwarf, first and second rows list values at $t=t_{1p}$ and $t=t_{2p}$, respectively.
    Note that, because of the nearly polar orbit of LMCa, the angular momentum 
    points in all cases in the $-X$ direction, i.e., to the Sun from the
    Galactic center. The angular momentum direction is also  listed
    for satellites with published proper motions (third row), and may be used to
    assess their possible association with the Clouds.}
\begin{tabular}[width=0.85\linewidth,clip]{|c|c|c|c|c|}
\hline
Name & time &  $j_X$ & $j_Y$ & $j_Z$ \\
\hline
\hline
LMC & $t=t_{1p}$ & $-0.97 \pm 0.03$ &  $0.14 \pm 0.07$ & $-0.19 \pm 0.10$\\ 
    & $t=t_{2p}$ & $-0.97 \pm  0.03$ & $0.14 \pm 0.06$ &  $-0.18 \pm 0.09$\\ 
    & obs        & $-0.97 \pm 0.01$ & $0.14 \pm 0.02$  &  $-0.18 \pm 0.03$\\
\hline 
SMC & $t=t_{1p}$ & $-0.92 \pm 0.05$ & $0.04 \pm 0.10$ & $-0.35 \pm 0.08$ \\ 
    & $t=t_{2p}$ & $-0.90 \pm 0.05$ & $0.05 \pm 0.17$ & $-0.38 \pm 0.10$  \\ 
    & obs & $-0.91 \pm 0.05$ & $0.08 \pm 0.11$ & $-0.39 \pm 0.09$ \\
\hline
Carina & $t=t_{1p}$ & $-0.93 \pm 0.12$ & $0.25 \pm 0.07$ & $-0.04 \pm 0.20$ \\
       & $t=t_{2p}$ & $-0.95 \pm 0.03$ & $0.25 \pm 0.06$ & $-0.01 \pm 0.15$ \\
       &    obs    & $-0.40  \pm 0.48$ & $0.46 \pm 0.18$ & $-0.79 \pm 0.38$\\
\hline
Fornax &  $t=t_{1p}$ & $-0.92 \pm 0.20$ & $0.19 \pm 0.11$ & $0.20 \pm 0.07$ \\
       &  $t=t_{2p}$ &  -- & -- & -- \\
       & obs & $-0.77  \pm 0.04$ & $0.63 \pm 0.05$ & $0.00 \pm 0.03$ \\
\hline
Sculptor &  $t=t_{1p}$ & -- & -- & -- \\ 
         &  $t=t_{2p}$ & $-0.94 \pm 0.06$ & $0.00 \pm 0.41$ & $0.04 \pm 0.05$ \\
         & obs &  $0.86 \pm 0.03$ & $-0.50 \pm 0.04$ & $0.003 \pm 0.005$  \\
\hline
Canes Venatici II &  $t=t_{1p}$ & -- & -- & -- \\
                  &  $t=t_{2p}$ & $-0.92 \pm 0.10$ & $-0.28 \pm 0.26$ & $-0.06 \pm 0.07$ \\
\hline
Leo II &  $t=t_{1p}$ & -- & -- & -- \\
       &  $t=t_{2p}$ & $-0.92 \pm 0.05$ & $0.21 \pm 0.21$ & $-0.28 \pm 0.15$ \\
\hline
Leo IV &  $t=t_{1p}$ & -- & -- & -- \\
       &  $t=t_{2p}$ & $-0.98 \pm 0.02$ & $0.16 \pm 0.12$ & $-0.01 \pm 0.09$ \\
\hline
Leo V  &  $t=t_{1p}$ & -- & -- & -- \\
       &  $t=t_{2p}$ & $-0.97 \pm 0.02$ & $0.17 \pm0.13$ & $-0.03 \pm 0.09$ \\
\hline
\hline
\end{tabular}
\label{TabPropMot1}
\end{table*}
\end{center}
%%%%%%%%%%%%%%%%%%%%

%TABLE%%%%%%%%%
\begin{center}
\begin{table*} 
  \caption{    List of properties of the Milky Way satellites used in Figs.~\ref{FigAitoff} and ~\ref{FigRVr}.
    The last three columns provide information about the likelihood of association of each dwarf with the LMC
    according to the sky proximity (Sec.~\ref{ssec:skyprox}), radial velocity (Sec.~\ref{ssec:vr}) and angular
    momentum (Sec.~\ref{ssec:angmom}) criteria. Pairs of symbols show the results for first and second pericenter (in that order).  A ``check-mark'' ($\checkmark$) indicates good or moderate agreement between
    the properties of a given dwarf and expectations from the simulated LMC analog. If, instead, there is a clear mismatch, the table
    displays ``$\times$''. 
    Question marks ($?$) in the last column denote the lack of available proper motions to compute the orbital angular momentum
    of these dwarfs.
    Previous association with the LMC requires the simultaneous fulfillment of the three conditions (see text for more detail).
    Note that the tests are applied recursively: the sky proximity is computed for all dwarf galaxies, but only those with a positive match
    are considered for the radial velocity and angular momentum tests. Therefore, after a ``$\times$'' symbol a dwarf is
    removed from the list of possible LMC companions, and a minus sign
    ``$-$'' fills the remaining columns. References are as follows:
    [1]: \citet{Mateo1998}, [2]: \citet{Simon_Geha2007}, [3]: \citet{Belokurov2010}, [4]: \citet{Belokurov2007},
    [5]: \citet{Belokurov2009}, [6]: \citet{Simon2007}, [7]: \citet{Martin2007}, [8]: \citet{Koch2009}, [9]: \citet{Irwin2007},
    [10]: \citet{Willman2005b}, [11]: \citet{Belokurov2008}.
  }
\begin{tabular}[width=0.85\linewidth,clip]{|c|c|c|c|c|c|c|c|c|c|}
\hline
Name & $M_v$ & $l$ & $b$ & $r$ & $V_r$ & Refs. &Test & Test & Test  \\
     & [mag] & $[^\circ]$ & $[^\circ]$ & kpc & km/s &  & Sky Proximity & Radial Velocity & Angular Momentum \\
\hline
\hline
LMC &  -18.5  &  280.5  &  -32.9  &   49.4  &   87.0  & [1] &  & & \\
SMC &  -17.1  &  302.8  &  -44.3  &   57.2  &   11.4  & [1] & $\checkmark$   $\checkmark$ & $\checkmark$   $\checkmark$ & $\checkmark$   $\checkmark$\\
Sculptor &   -9.8  &  287.5  &  -83.2  &   90.1  &   78.1 & [1]  & $\checkmark$   $\checkmark$ & $\times$   $\checkmark$ & $-$ $\times$ \\
Fornax &  -13.1  &  237.1  &  -65.7  &  142.1  &  -33.7 & [1]  & $\checkmark$   $\checkmark$ & $\checkmark$  $\checkmark$ & $\times$ $\times$ \\
Carina &   -9.4  &  260.1  &  -22.2  &  101.7  &   13.3 & [1]  & $\checkmark$  $\checkmark$ & $\checkmark$ $\checkmark$ & $\times$ $\times$ \\ 
Leo I &  -11.9  &  226.0  &   49.1  &  254.0  &  181.9 & [1]  & $\times$ $\checkmark$ & $-$ $\times$ & $-$ $-$\\ 
Sextans &   -9.5  &  243.5  &   42.3  &   93.2  &   78.6 & [1] & $\checkmark$ $\checkmark$ & $\times$ $\times$ & $-$ $-$\\ 
Leo II&  -10.1  &  220.2  &   67.2  &  212.7  &   23.4 & [1]  & $\times$ $\checkmark$ & $-$ $\checkmark$ & $- \;$ ? \\ 
Ursa Minor&   -8.9  &  105.0  &   44.8  &   62.1  &  -89.8 & [1]  & $\times$ $\times$ & $-$ $-$ & $-$ $-$\\ 
Draco &   -8.6  &   86.4  &   34.7  &   80.0  & -103.8 & [1]  & $\checkmark$ $\times$ & $\times$ $-$ & $-$ $-$ \\
Sagittarius&  -13.8  &    5.6  &  -14.1  &   21.9  &  171.9 & [1]  & $\times$ $\times$ &  $-$ $-$ & $-$ $-$\\ 
Tucana &  -10.0  &  322.9  &  -47.3  &  475.5  &   38.0 & [1]  & $\checkmark$ $\times$ & $\times$ $-$ & $-$ $-$ \\
Ursa Major I &   -5.6  &  159.4  &   54.4  &  110.9  &  -8.8 & [2]  & $\times$ $\times$ & $-$ $-$ & $-$ $-$ \\
Ursa Major II &   -3.8  &  152.4  &   37.4  &   38.5  & -36.6  & [2] &$\times$ $\times$ & $-$ $-$ & $-$ $-$ \\
Canes Venatici I &   -7.9  &   74.3  &   79.8  &  219.8  &   76.9  & [2] &  $\times$ $\checkmark$ & $-$ $\times$ & $-$ $-$ \\
Canes Venatici II &   -4.8  &  113.6  &   82.7  &  151.7  &  -96.2  & [2] & $\times$ $\checkmark$ & $-$ $\checkmark$ & $- \;$ ? \\
Pisces II &   -5.0  &   79.2  &  -47.1  &    $-$  &  202.4  & [3] & $\times$ $\times$ & $-$ $-$ & $-$ $-$\\
Segue 1 &   -3.0  &  220.5  &   50.4  &   28.1  &  116.3  & [2][4] & $\times$ $\times$ & $-$ $-$ & $-$ $-$\\
Segue 2 &   -2.5  &  149.4  &  -38.1  &   41.2  &   40.1  & [5] & $\times$ $\times$ & $-$ $-$ & $-$ $-$ \\
Segue 3 &   -1.2  &   69.4  &  -21.2  &    $-$  &  209.7  & [3] & $\times$ $\checkmark$ & $-$ $\times$ & $-$ $-$\\
Coma &   -3.7  &  241.9  &   83.6  &   45.2  &   83.9  & [4][6] & $\times$ $\checkmark$ & $-$ $\times$ &  $-$ $-$ \\
Hercules &   -6.0  &   28.7  &   36.9  &  132.2  &  142.9 & [4] & $\times$ $\times$ & $-$ $-$ & $-$ $-$ \\
Leo IV &   -5.1  &  265.4  &   56.5  &  158.6  &   14.0  & [4] & $\times$ $\checkmark$ & $-$ $\checkmark$ & $- \;$ ? \\
Bootes &   -5.8  &  358.1  &   69.6  &   63.5  &  107.4  & [7] & $\times$ $\checkmark$ & $-$ $\times$ & $-$ $-$ \\
Bootes II&   -2.7  &  353.7  &   68.8  &   44.7  & -117.3  & [8] &$\times$ $\checkmark$ & $-$ $\times$ & $-$ $-$\\
Leo T &   -7.1  &  214.8  &   43.6  &  422.1  &  -56.0  & [9] & $\times$ $\checkmark$ & $-$ $\times$ & $-$ $-$\\
Willman 1&   -2.7  &  158.5  &   56.7  &  484.4  &   34.2  & [10] & $\times$ $\checkmark$ & $-$ $\times$ & $-$ $-$\\
Leo V &   -4.3  &  261.9  &   58.5  &  180.8  &   62.3  & [11] &$\times$ $\checkmark$ & $-$ $\checkmark$ & $- \;$ ?\\
\hline
\hline
\end{tabular}
\label{TabMWsats}
\end{table*}
\end{center}
%%%%%%%%%%%%%%%%%%%%

\section*{Acknowledgements}
\label{acknowledgements}

We would like to thank Volker Springel and Adrian Jenkins for their 
important contributions to the simulations on which this work is based.
LVS is grateful for financial support from the {\it CosmoComp/Marie Curie} network. 
CSF acknowledges a Royal Society Wolfson research merit award and ERC
Advanced Investigator grant COSMIWAY. This work was supported in part by
an STFC rolling grant to the ICC.
The simulations for the Aquarius Project were carried out at the
Leibniz Computing Centre, Garching, Germany, at the Computing
Centre of the Max-Planck-Society in Garching, at the Institute for
Computational Cosmology in Durham, and on the ‘STELLA’ supercomputer of
the LOFAR experiment at the University of Groningen. We thank the
anonymous referee for a very careful and useful report.

\bibliography{master}

%%%%%%%%%%%%%%%%%%%%%%%%%%%%%%%%%%%%%%%%%%%%%%%%
%%  APPENDIX    %%
%%%%%%%%%%%%%%%%%%%%%%%%%%%%%%%%%%%%%%%%%%%%%%%%

%\appendix{{\it Magellanic Galaxy}-debris in the LMC frame}
\appendix
\section{{\it Magellanic Galaxy}-debris in the LMC frame}

\begin{center} \begin{figure*} 
\includegraphics[width=0.475\linewidth,clip]{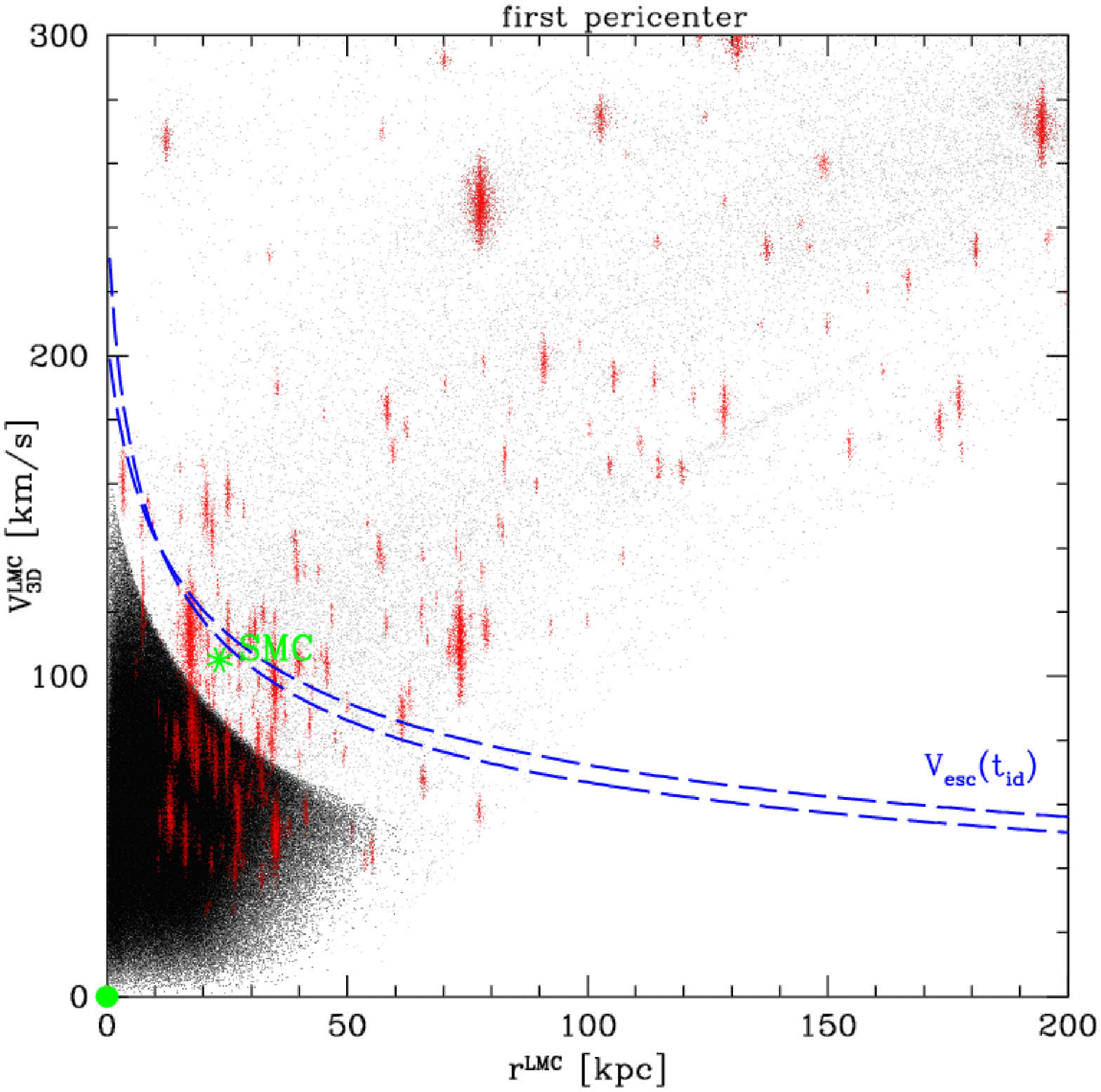} 
\includegraphics[width=0.475\linewidth,clip]{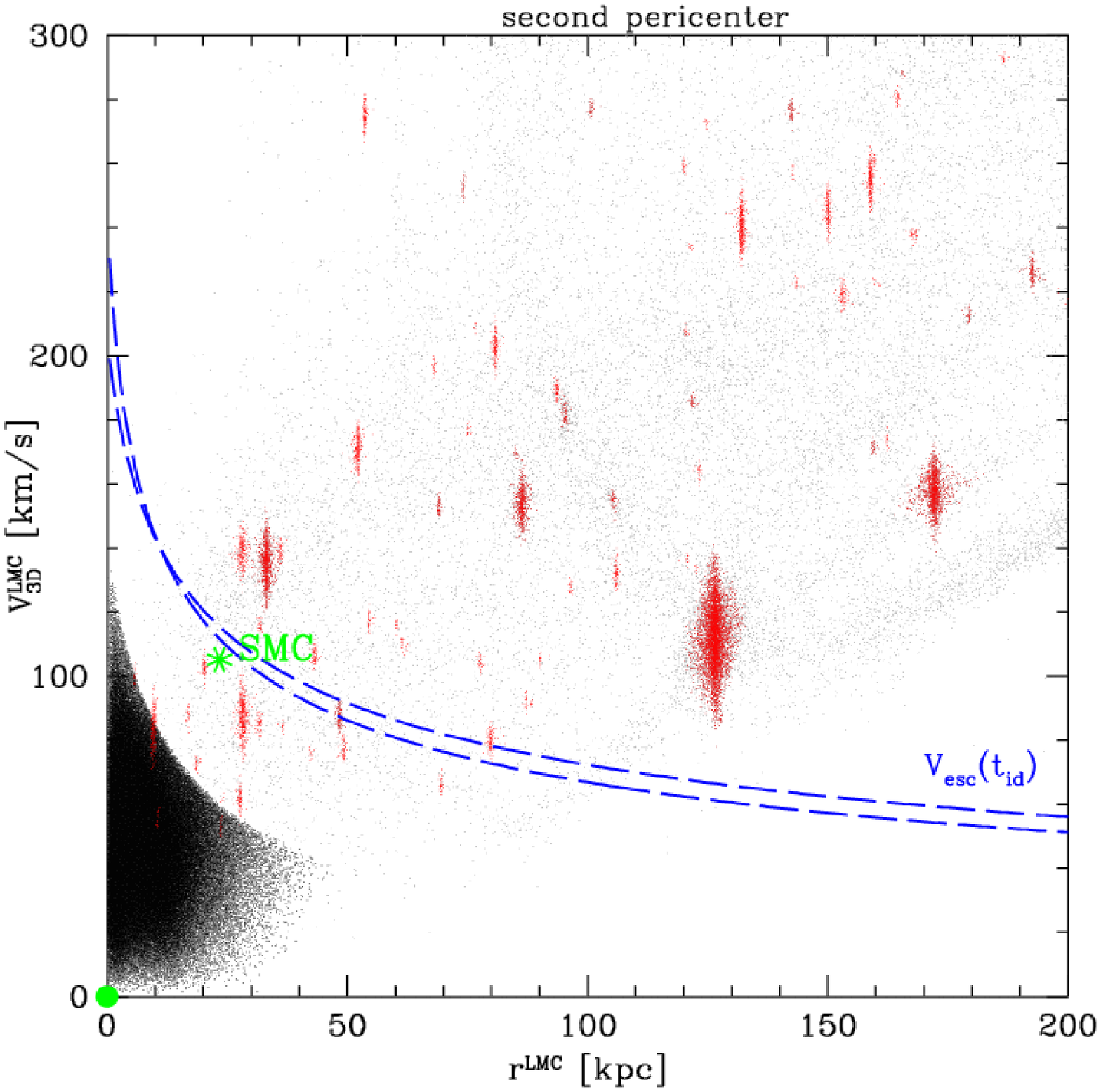} 
\caption{Total 3D velocity, $V^{LMC}_{3D}$, versus distance $r^{LMC}$,
  where upperscripts indicate that they are computed with respect to
  the LMC center. Color coding is the same as in the previous
  figure. The green asterisk shows the position of the SMC in this
  plane according to data from Kallivayalil et al. 2006b.  The
  transition from black (bound) to gray (unbound) dots can be used to
  infer the instantaneous escape velocity of the LMCa system. For
  comparison, blue curves show $V_{esc}$ for an NFW halo with the mass
  of LMCa at the time of infall, $M_{200}=3.6 \times 10^{10} M_\odot$,
  assuming two different concentrations c=10,20.  The effect of tides
  due to the host potential can be seen from the comparison between
  these curves and the velocity of the bound (black) particles; in
  particular, the mass loss experienced between first and second
  approach is reflected by the smaller area (and lower velocities)
  covered by black particles in the right panel (second pericenter)
  compared to the left (first pericenter).   
  }
\label{fig_appendix}
\end{figure*}
\end{center}

For completeness, Fig.~\ref{fig_appendix} shows the distance-velocity
plane of LMCa debris in a coordinate system centered at the LMC. As
the system orbits within the host potential, mass is lost to tides and
the material initially associated to the LMC group gets progresively
unbound. Substructures that remain bound to the LMC are shown in
black, unbound material in grey. LMCa substructures are shown in
red. The SMC falls in the ``unbound'' region at either first or second
pericenter, which suggests that the LMC-SMC might not longer be a
bound pair. This conclusion, however, is sensitive to the mass of LMCa
which, as discussed in the text, is likely smaller than the true LMC
mass. Once reliable three-dimensional velocities become available for
more dwarf galaxies, their distribution in the LMC-centered phase
space may be used to place further constraints on their association to
the Clouds.

%%%%%%%%%%%%%%%%%%%%%%%%%%
%%%%%%%%%%%%%%%%%%

\end{document}